\documentclass[9pt,twocolumn,twoside]{osajnl}

\journal{ol} 

\setboolean{shortarticle}{true}

\title{Optomechanical quadrature squeezing in the non-Markovian regime}

\author[1]{Biao Xiong}
\author[1]{Xun Li}
\author[1]{Shi-Lei Chao}
\author[*]{Ling Zhou}

\affil[1]{School of Physics, Dalian University of
Technology, Dalian 116024,People's Republic of China}

\affil[*]{Corresponding author: zhlhxn@dlut.edu.cn}

\begin{abstract}
Squeezing of quantum fluctuation plays an important role in fundamental quantum physics and has marked influence on ultrasensitive detection. We propose a scheme to generate and enhance the squeezing of mechanical mode by exposing the optomechanical system to a non-Markovian environment. It is shown that the effective parametric resonance term of mechanical mode can be induced due to the interaction with cavity and non-Markovian reservoir, thus resulting in quadrature squeezing of the mechanical resonator. And jointing the two kinds of interactions can enhance the squeezing effect. Comparing with the usual Markovian regime, we can obtain stronger squeezing, and significantly the squeezing can approach a low asymptotic stable value.
\end{abstract}

\setboolean{displaycopyright}{true}

\begin{document}

\maketitle

Quantum physics exhibits many interesting non-calssical effects \cite{PhysRevA.82.045805,6046093,Berrada2013,Eleuch2008,Eleuch2010}. Quantum fluctuations, originated from the Heisenberg uncertainty principle, are the unique properties of quantum physics. Several well known interesting physical phenomena such as Casimir forces and the Lamb shift \cite{Clark2017Sideband} are produced by quantum fluctuations. While quantum fluctuations also result in some restrictions in precision measurements. For example, the vacuum fluctuations in the optomechanical systems can broaden the optical response spectrum and affect the sensitivity of detection \cite{PhysRevLett.115.243603,PhysRevLett.113.151102,1367-2630-19-8-083022}. Fortunately, vacuum fluctuations are not immutable and can be ‘squeezed’ \cite{PhysRevLett.118.143601,Liu:18,Liu:16}. For a squeezed state, the quantum fluctuations in one variable are reduced below their value at the expense of the corresponding increased fluctuation in the conjugate variable \cite{Scully1998Quantum}. By using the squeezing state, the precision of position measurement can be beyond the standard quantum limit \cite{1367-2630-18-7-073040}. Therefore, squeezed states of light are useful in high precision measurements such as gravitational wave detection \cite{Abadie2011A,PhysRevLett.110.181101}.

Optomechanical system is a hot topic in recent years \cite{PhysRevA.97.023841,PhysRevA.96.042320,PhysRevA.96.021801}. The squeezing of mechanical modes can be utilized to improve the sensitivity of precision quantum measurement in cavity optomechanical systems \cite{RevModPhys.86.1391}. Enormous schemes and methods of squeezing mechanical mode have been theoretically proposed and experimentally realized in optomechanical system \cite{PhysRevLett.57.2520,PhysRevLett.76.2294,doi:10.1021/nl101844r,PhysRevLett.98.078103,PhysRevA.91.013834,Wang2016Steady2,PhysRevA.89.023849,PhysRevA.89.063805,PhysRevA.87.033829,PhysRevA.88.063833,PhysRevA.83.033820,Han2013}. The basic mechanism for preparing quadrature squeezing of mechanical oscillator is to introduce an effective mechanical parametric amplification \cite{PhysRevLett.57.2520,PhysRevLett.76.2294,doi:10.1021/nl101844r}. This can be realized by utilizing the intrinsic nonlinearity of mechanical oscillator \cite{PhysRevLett.98.078103,PhysRevA.91.013834,Wang2016Steady2}, employing impulse kicks on a mechanical oscillator \cite{PhysRevA.89.023849} and engineering the quantum reservoir \cite{PhysRevA.89.063805,PhysRevA.87.033829,PhysRevA.88.063833}, etc. In Ref.~\cite{PhysRevA.83.033820,Han2013}, the mechanical squeezing was generated by modulating the driving field at a frequency matching to the frequency shift of the mirror under the condition of large frequency detuning, thus approximately reaching the parametric resonance. However, the squeezing is oscillated with time evolution, which make it difficult for applications.

In cavity optomechanical system, when linearization of cavity field is performed, the interaction between cavity field and mechanical mode is of $XX$ (coordinate-coordinate) type, therefore the parametric resonance term of mechanical mode is produced after eliminating the cavity mode, therefore the squeezing effect is obtained. Similarly, the $XX$ interaction between mechanical oscillator and its environment should also contribute to squeezing effect, while the non-Markovian environment might benefit to keep the effect of $XX$ type interaction. Recently, the features of non-Markovian process have sparked a great interest in theoretical study \cite{PhysRevA.91.022328,PhysRevA.94.012334,PhysRevD.45.2843,PhysRevA.91.012109,PhysRevLett.109.170402,RevModPhys.88.021002,PhysRevA.96.062114,PhysRevA.93.063853,Shen:18,Valente:16,Li:17,PhysRevA.97.012104,LI20182044,1367-2630-20-5-053026,RevModPhys.89.015001}, and the non-Markovian spectral density has been detected in a micro-optomechanical system \cite{groblacher2015observation}, which make it necessary to research the squeezing effect of optomechanical system in the non-Markovian environment. 

\begin{figure}
\centering
  \includegraphics[width=3.0in]{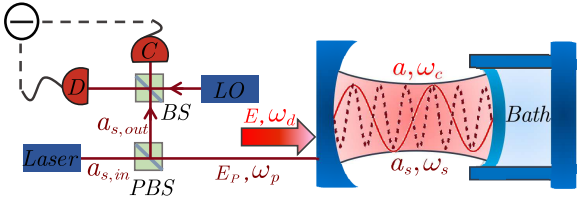}
  \caption{Schematic of the squeezing generation and detection in cavity-optomechanical system, where the mechanical oscillator couples to a general non-Markovian reservoir. 
  }\label{Fig1}
\end{figure}

As shown in Fig.~\ref{Fig1}, we consider a generic cavity optomechanical system, where the cavity field with frequency $\omega_{c}$ couples to a mechanical oscillator with frequency $\omega_{m} $ with the coupling strength $g_{0}$ due to the radiation pressure. The mechanical oscillator is surrounded by a non-Markovian reservoir. The Hamiltonian of the system can be written as $%
H=H_{S}+H_{E}+H_{I}$, where
\begin{subequations}
\begin{align}
H_{S} &=\hbar \omega _{c}a^{\dag }a+\hbar \omega _{m}b^{\dag }b-\hbar
g_{0}a^{\dag }a(b^{\dag }+b)  \notag \\
&+i\hbar E(t)(a^{\dag }e^{-i\omega _{d}t}-ae^{i\omega _{d}t}),
\label{HS} \\
H_{E} &=\sum_{k}\hbar \omega _{k}b_{k}^{\dag }b_{k}, \quad H_{I} = \sum_{k}\hbar V_{k}(b+b^{\dag })(b_{k}^{\dag }+b_{k}). \label{HI}
\end{align}
\end{subequations}
Here $H_{s}$ describes a normal optomechanical system driven by a classical laser with the frequency $\omega_{d}$ and driven strength $E$. $H_{E}$ is the free energy of the mechanical reservoir with the frequency  $\omega_{k}$ of the $k$th bath mode. $H_{I}$ denotes the $XX$ type interaction between the mechanical oscillator and the reservoir, with $V_{k}$ the coupling strength. The $XX$ type non-Markovian interaction between the micro-mechanical resonator and its bath can be realized experimentally with a high-reflectivity mirror pad in the center of a $Si_{3}N_{4}$ beam in vacuum \cite{groblacher2015observation}. 
In the frame rotating with $H_{0}=\omega_{d}a^{\dagger}a$, the Langevin equations are given by
\begin{subequations}
\begin{align}
\dot{a} &=-(i\Delta _{c}+\frac{\kappa }{2})a+ig_{0}a(b+b^{\dag })+E(t)+\sqrt{
\kappa }a_{\text{in}}, \label{Langevin_a} \\
\dot{b} &=-i\omega _{m}b+ig_{0}a^{\dag }a -i\sum_{k}V_{k}(b_{k}+b_{k}^{\dag
}), \label{Langevin_b} \\
\dot{b}_{k} &=-i\omega _{k}b_{k}-iV_{k}(b+b^{\dag }),  \label{Langevin_bk}
\end{align}
\end{subequations}
where $\Delta _{c}=\omega _{c}-\omega _{d}$, and $a_{\text{in}}$ is the input
noise operator of the cavity. Solving Eq.~(\ref{Langevin_bk}) for the bath
operator $b_{k}(t)$ and substituting the formal solution of $b_{k}(t)$ into Eq.~(\ref{Langevin_b}), we obtain
\begin{align}   \label{Formal_Solution_b}
\dot{b} =-i\omega _{m}b+ig_{0}a^{\dag }a +\int_{0}^{t}d\tau f(t-\tau )[b(\tau )+b^{\dag }(\tau )]-\xi (t). 
\end{align}
The formal integral in Eq.~(\ref{Formal_Solution_b}) describes the backflow from the non-Markovian environment, where the memory kernel $f(t)=2i\sum_{k}V_{k}^{2}\sin (\omega _{k}t)=2i\int_{0}^{\infty
}d\omega \mathcal{J}(\omega )\sin (\omega t)$ with $\mathcal{J}(\omega )$
the spectral density of the reservoir, describing the structure of the reservoir, as well as the coupling between the reservoir and the system. The commonly used spectral density $\mathcal{J}(\omega )=\eta \omega (\omega /\omega_{0})^{s-1}e^{-\omega /\omega _{0}}$~\cite{RevModPhys.59.1} is employed in this paper, where $\eta $ is the strength of the system-bath coupling and $\omega _{0}$ is the cut-off frequency. The exponent $s$ is a real number with $0<s<1$, $s=1$, and $s>1$ classifying the bath as sub-Ohmic, Ohmic, and super-Ohmic, respectively. The noise
term $\xi (t)=i\sum_{k}V_{k}[b_{k}(0)e^{-i\omega _{k}t}+b_{k}^{\dag}(0)e^{i\omega _{k}t}]$
in Eq.~(\ref{Formal_Solution_b}) is determined by the initial states of the environment. Instead of $\langle \xi(t)\xi(t^{\prime}) \rangle \varpropto \delta (t-t^{\prime })$ for the Markovian environment, the noise correlation function is 
\begin{eqnarray}
\langle \xi(\tau_{1})\xi(\tau_{2}) \rangle &=& -\int_{0}^{\infty }\mathcal{J}(\omega )d\omega
\{e^{-i\omega (\tau _{1}-\tau _{2})} \notag\\
&&+2\cos \omega (\tau _{1}-\tau _{2})(e^{\frac{\hbar \omega }{k_{B}T}%
}-1)^{-1}\},
\end{eqnarray}
where we assume  the initial correlation function of the reservoir as $\langle b_{k}^{\dag }(0)b_{k}(0)\rangle =m_{k}$ with $m_{k}=1/(e^{\hbar\omega _{k}/k_{B}T}-1)$ the distribution function of the reservoir. 

Eqs.~(\ref{Langevin_a}) and (\ref{Formal_Solution_b}) can be linearized by $a\rightarrow \alpha+\delta a$ and
$b \rightarrow \beta+\delta b$. Here $\alpha$ and $\beta$ are $C-$numbers denoting the mean values of the optical and mechanical modes. $\delta a$ and $\delta b$ describe the quantum fluctuations of optical and mechanical modes. For simplicity, the symbol $\delta$ in $\delta a$ and $\delta b$ are neglected in the remind part of this paper. Then the motion equations satisfy
\begin{subequations} \label{LLangevin}
\begin{align}
\dot{\alpha} =&-(i\Delta _{c}+\frac{\kappa }{2})\alpha +ig_{0}\alpha (\beta
+\beta ^{\ast })+E(t),  \label{LL_alpha} \\
\dot{\beta} =&-i\omega _{m}\beta +ig_{0}|\alpha |^{2} +\int_{0}^{t}d\tau f(t-\tau )[\beta (\tau )+\beta ^{\ast }(\tau )],
\label{LL_belta} \\
\dot{a} =&-(i\Delta _{c}^{\prime }+\frac{\kappa }{2})a+iG(b+b^{\dag })+
\sqrt{\kappa }a_{\text{in}},  \label{LL_a} \\
\dot{b} =&-i\omega _{m}b+i(Ga^{\dag }+G^{\ast }a) +\int_{0}^{t}d\tau f(t-\tau )[b(\tau )+b^{\dag }(\tau )] \notag \\
&-\xi (t),
\label{LL_b}
\end{align}
\end{subequations}
where $ \Delta_{c}^{\prime} = \Delta_{c}-g_{0}(\beta+\beta^{*})$ is the effective detuning and $G = \alpha g_{0}$ describes the effective linear coupling strength. 

In order to study the dynamics of mechanical oscillator, we first solve Eq.~(\ref{LL_a}) and then substitute the solution into Eq.~(\ref{LL_b}), we finally get
\begin{eqnarray}  \label{S_b} 
\dot{b} &=&-i \omega _{m}b+\int_{0}^{t}d\tau F(t-\tau )[b(\tau )+b^{\dag
}(\tau )]+S_{in}(t),
\end{eqnarray}
where
\begin{subequations}
\begin{align} \label{S_b_Supp}
F(t-\tau ) =f(t-\tau )-[G^{\ast }(t)G(\tau )e^{u(t-\tau )}-H.c.], \\
S_{in}(t) =A_{0}(t)+A_{\text{in}}(t)-\xi (t), \\
A_{0}(t) =i[G^{\ast }(t)e^{u(t)}a(0)+H.c.],  \\ 
A_{\text{in}}(t) =\int_{0}^{t}d\tau i[\sqrt{\kappa }G^{\ast
}(t)e^{u(t-\tau )}a_{\text{in}}(\tau )+H.c.],  \\
u(t_{1}-t_{2})=-\int_{t_{2}}^{t_{1}}d\tau \lbrack i\Delta ^{\prime
}(\tau )+\kappa /2].
\end{align}
\end{subequations}
$F(t-\tau)$ is the memory kernel resulting from cavity and reservoir. It contains two parts. The one is the backflow from the non-Markovian reservoir, and the other results from the interaction between the cavity field and mechanical mode. We will show that the two interactions will both contribute to the squeezing of mechanical oscillator. $A_{0}$ and $A_{in}(t)$ reflect the impact of initial condition and thermal noise of cavity field, respectively. 
To solve Eq.~(\ref{S_b}), we assume
\begin{equation}  \label{Assume_b}
b(t)=M(t)b(0)+L^{\ast }(t)b^{\dag }(0)+S(t), 
\end{equation}
where $M(t)$ and $L(t)$ are complex numbers with the initial conditions $M(0)=1$, $L(0)=0$, and $S(t)$ is a operator with $S(0)=0$. By substituting Eq.~(\ref{Assume_b}) into
Eq.~(\ref{S_b}), we obtain
\begin{subequations} \label{MLS_Equation}
\begin{align} 
\dot{M}(t) &=-i\omega _{m}M(t)+\int_{0}^{t}d\tau F(t-\tau )[M({\tau })+L(\tau )],     \\
\dot{L}(t) &=i\omega _{m}L(t)+\int_{0}^{t}d\tau F^{\ast }(t-\tau )[M({\tau })+L(\tau )],   \\
\dot{S}(t) &=-i\omega _{m}S(t)+\int_{0}^{t}d\tau F(t-\tau )[S(\tau)+S^{\dag }(\tau )]  +S_{in}.
\end{align} 
\end{subequations} 
If $M(t)$ and $L(t)$ are known, the operator $S(t)$
can be completely determined through \cite{PhysRevA.81.052105}
\begin{align} \label{S_Equation}
S(t) = \int_{0}^{t}d\tau \lbrack M(t-\tau )-L^{\ast }(t-\tau )] S_{in}(\tau).
\end{align}
Therefore, the quadrature fluctuation of mechanical mode $\Delta X(\theta,t)=\langle X(\theta,t)^{2} \rangle-\langle X(\theta,t)\rangle^{2}$ can be solved numerically since $b$ is determined by Eq.~(\ref{Assume_b}), (\ref{MLS_Equation}) and (\ref{S_Equation}), where $X(\theta,t)=X(t)\cos \theta+Y(t)\sin \theta$ is the quadrature operator with $X(t)=[b(t)+b^{\dagger}(t)]/\sqrt{2}$ and $Y(t)=-i[b(t)-b^{\dagger}(t)]/\sqrt{2}$.

In order to clearly understand the process of quadrature squeezing generation, we rewrite Eq.~(\ref{S_b}) as
\begin{align}  \label{S_b_rewrite} 
\dot{b} =&-i \omega _{m}b+\int_{0}^{t}d\tau F(t-\tau )b(\tau )+\int_{0}^{t}d\tau F(t-\tau ) b^{\dag}(\tau ) + S_{in}(t).
\end{align}
Here the first integration in the right hand side of Eq.~(\ref{S_b_rewrite}) is the free term of mechanical oscillator induced by the cavity and reservoir, which can modulate the free frequency of mechanical oscillator. The second integration is the nonlinear degenerate parametric resonance term which may result in the squeezing of mechanical oscillator. In the Markovian regime the  dynamic equation becomes
\begin{align}  \label{S_b_rewrite_M} 
\dot{b} =& -(i \omega _{m}+\gamma/2)b-\int_{0}^{t}d\tau [G^{\ast }(t)G(\tau )e^{u(t-\tau )}-H.c.] b^{\dag}(\tau ) \notag \\
& +A_{0}(t)+A_{\text{in}}(t)+\sqrt{\gamma}b_{in}.
\end{align}
From Eq.~(\ref{S_b_rewrite_M}), we note that the parametric resonance term of mechanical mode  is now induced merely by the interaction with the cavity field, which may result less squeezing than that in the non-Markovian case.
\begin{figure}[htbp]
\centering
  \includegraphics[width=\linewidth]{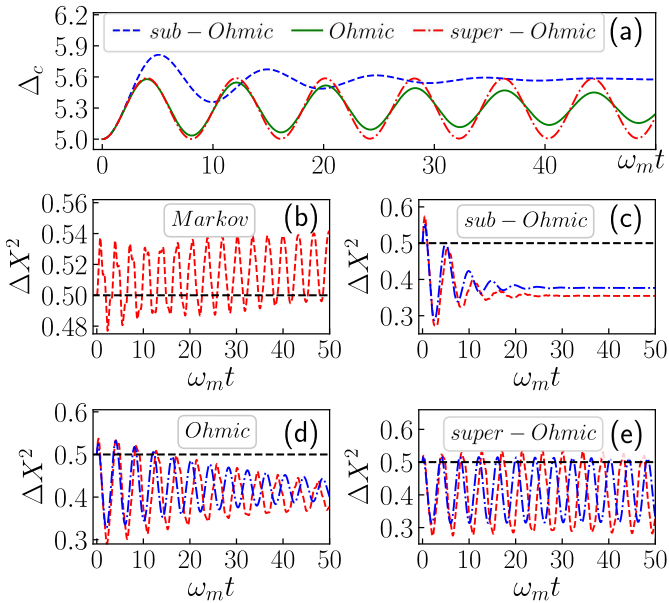}
  \caption{(Color online) (a) The modulated detuning $\Delta_{c}$ to ensure the fixed $\Delta_{c}^{\prime}=5\omega_{m}$. (b)-(e) Time evolution of quadrature squeezing $\Delta X^{2}$ with $G=0$ (dashed-dotted blue line) and $G=0.3$ (dashed red line) at fixed $\Delta_{c}^{\prime}=5\omega_{m}$ in different regime. Here $\gamma=1\times 10^{-9}\omega_{m}$ for Markovian case, $s=0.5$, $1$ and $2$ for sub-Ohmic, Ohmic and super-Ohmic, respectively. The other parameters are $g=10^{-4}$, $\theta=\pi/2$, $T=0$, $\omega_{0}=20\omega_{m}$, $\eta=5\times10^{-3}$ and $\kappa=0.1\omega_{m}$.
  }\label{Fig2}
\end{figure}

To verify our analysis, we simulate the evolution of quadrature fluctuation at fixed $\Delta_{c}^{\prime}=5\omega_{m}$ in Fig.~\ref{Fig2} (b)-(e). The values of $\kappa$, $\gamma$ and $g$ taken in Fig.~\ref{Fig2} are within the scopes of current experiments summarized in  \cite{RevModPhys.86.1391}. In principle, $\Delta_{c}^{\prime}$ and $G$ determined by Eq.~(\ref{LL_alpha}) and (\ref{LL_belta}) are time-dependent. Actually, the controllable parameters $\Delta_{c}$ and $E(t)$ can ensure that $\Delta_{c}^{\prime}$ and $G$ get desired values \cite{PhysRevLett.116.183602}. In the following simulation, we set $\Delta_{c}^{\prime}$ and $G$ time-independent. To ensure the fixed $\Delta_{c}^{\prime}=5\omega_{m}$, we simulate the evolution of $\Delta_{c}$ in Fig.~{\ref{Fig2}} (a). It is shown that $\Delta_{c}$ varies with time continuously, which obviously can be realized by the frequency modulation method, thus the time-independent value of $\Delta_{c}^{\prime}$ is achievable. In Fig.~{\ref{Fig2}} (b) the extremely small dissipation $\gamma=10^{-9}$ are considered in the usual Markovian regime. We see that the squeezing is weak, and will disappear in the longer time in Markovian regime. Because of the effective beam-splitter interaction between the mechanical mode and its bath in the usual Markovian regime, the bath of mechanical oscillator can't induce the squeezing of mechanical mode. The squeezing in  Fig.~\ref{Fig2} (b) completely results from the $XX$ type interaction between the mechanical mode and cavity field as Eq.~(\ref{S_b_rewrite_M}) shown. In Fig.~\ref{Fig2} (c)-(e), we simulate the quadrature fluctuation in the non-Markovian regime with $G=0$ and $G=0.3$, respectively. Comparing the dashed red line of \ref{Fig2} (c)-(e) with Fig.~\ref{Fig2} (b), we see that the stronger squeezing can be achieved in the non-Markovian regime. We can understand this easily because the $XX$ type non-Markovian interaction between mechanical mode and its bath can also induce the squeezing of mechanical mode, as the first part of $F(t-\tau)$ in Eq.~(\ref{S_b_Supp}) shown. Significantly, the squeezing can achieve asymptotic stable for the sub-Ohmic bath within the time interval shown in Fig.~\ref{Fig2} (b), which is meaningful to experiments.

The dashed-dotted blue line in Fig.~\ref{Fig2} (c)-(e) is corresponding to $G=0$, which means that the mechanical oscillator is in the free space, in which the squeezing is completely induced by the bath. Comparing the dashed-dotted blue line ($G=0$) with the dashed red line ($G=0.3$) in Fig.~\ref{Fig2} (c)-(e) respectively, we see that the squeezing with $G=0.3$ is stronger than that with $G=0$. This phenomenon can be explained by Eq.~(\ref{S_b_rewrite}), since the coefficient $F(t-\tau)$ in the second integration consists of both the cavity and bath induced part. Therefore the joint effect of the two kinds of interactions is better than that with only one.
\begin{figure}[htbp]
\centering
  \includegraphics[width=\linewidth]{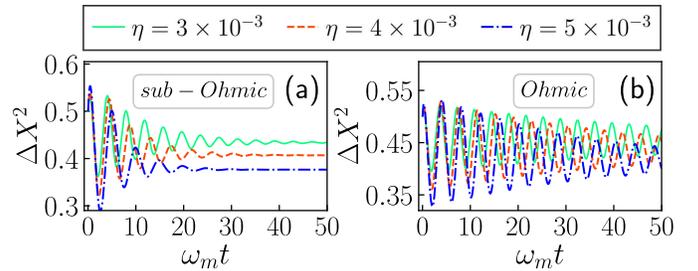}
  \caption{(Color online) Time evolution of quadrature squeezing $\Delta X^{2}$ with different $\eta$ at fixed $G=0$ in sub-Ohmic (a) and Ohmic (b) regime. The other parameters are the same as in Fig.~\ref{Fig2}.}\label{Fig3}
\end{figure}

Now, we further show the effect of the parameter $\eta$ on the squeezing of mechanical mode. The evolution of fluctuation with different $\eta$ in sub-Ohmic regime and Ohmic regime are simulated in Fig.~\ref{Fig3}(a) and (b), respectively. We see that the larger $\eta$, the stronger squeezing. Since $\eta$ describes the coupling strength between the mechanical oscillator and its bath, as the second integration shown in Eq.~(\ref{S_b_rewrite}), the larger $\eta$, the stronger backaction of the bath in mechanical oscillator, thus inducing stronger squeezing. Therefore, to obtain strong squeezing, the coupling strength between the mechanical mode and environment should be large in practical applications.

We finally discuss the experimental detection of the generated quadrature squeezing of the mechanical mode, which could be realized by introducing an ancillary cavity mode. As depicted in Fig.~\ref{Fig1}, the ancillary cavity mode $a_{s}$ with resonant frequency $\omega_{s}$ is driven by a pump field with amplitude $E_{p}$ and frequency $\omega_{p}$. If the driven $E_{p}$ is very weak so that the effective coupling $G_{s}\ll G$, the backaction of the ancillary cavity mode on the mechanical mode can be neglected and the dynamic of the system is still well described by Eqs.~(\ref{LLangevin}) \cite{PhysRevA.91.013834}. In the interaction picture, by choosing the effective detuning $\Delta_{s}^{'}=\omega_{m}$ and neglecting the rapid oscillating term, we have
\begin{eqnarray}
\dot{a_{s}}&=&-\frac{\kappa_{s}}{2} a_{s} + iG_{s}b+\sqrt{\kappa_{s}}a_{s,in},
\end{eqnarray}
where $\kappa_{s}$ is the dissipation of $a_{s}$. If $\kappa_{s} \gg G_{s}$, the ancillary cavity mode adiabatically follows the dynamics of mechanical mode, thus leading 
\begin{eqnarray}
a_{s,out}&=&\frac{2iG_{s}}{\sqrt{\kappa_{s}}} b + a_{s,in},
\end{eqnarray}
where the input-output relation $a_{s,out}=\sqrt{\kappa_{s}}a_{s}-a_{s,in}$ has been used. We see that the output field of $a_{s,out}$ gives a direct measurement of mechanical mode $b$, therefore the squeezing of mechanical mode can be detected by directly homodyne output field $a_{s,out}$  with a local oscillator (LO). The balanced homodyne detection method can be found in \cite{Scully1998Quantum}, we will not repeat here.

In this paper, a scheme of quadrature squeezing of the mechanical oscillator for cavity optomechanical system has been investigated. By introducing the $XX$ type interaction between the oscillator and its bath, we find that the bath can induce the squeezing as the optomechanical interaction can do. For the cavity optomechanical system, the asymptotic squeezing after long time evolution can be enhanced by considering the non-Markovian effect. Moreover, our scheme can generate asymptotic squeezing without assistant parametric resonance. The parametric resonance term of mechanical mode in our scheme is completely induced by the optomechanical radiation pressure coupling and oscillator-reservoir coupling. Although the squeezing in Refs~\cite{PhysRevA.83.033820,Han2013} is obtained without assistant parametric amplification, the squeezing oscillates with time evolution, which is difficult to the application. Our scheme obtaining asymptotic squeezing is of practical significance. Our scheme of the non-Markovian bath can be realized in experiment\cite{groblacher2015observation}, therefore, the scheme is feasible.

\section{Funding Information}
National Science Foundation of China (NSFC) (11474044, 11874099).

\begin{thebibliography}{10}
\newcommand{\enquote}[1]{``#1''}

\bibitem{PhysRevA.82.045805}
P.~K. Jha, H.~Eleuch, and Y.~V. Rostovtsev, {\protect\JournalTitle{Phys. Rev.
  A}} \textbf{82}, 045805 (2010).

\bibitem{6046093}
E.~A. Sete, A.~A. Svidzinsky, Y.~V. Rostovtsev, H.~Eleuch, P.~K. Jha,
  S.~Suckewer, and M.~O. Scully, {\protect\JournalTitle{IEEE Journal of
  Selected Topics in Quantum Electronics}} \textbf{18}, 541 (2012).

\bibitem{Berrada2013}
K.~Berrada, S.~Abdel-Khalek, H.~Eleuch, and Y.~Hassouni,
  {\protect\JournalTitle{Quantum Information Processing}} \textbf{12}, 69
  (2013).

\bibitem{Eleuch2008}
H.~Eleuch, {\protect\JournalTitle{The European Physical Journal D}}
  \textbf{49}, 391 (2008).

\bibitem{Eleuch2010}
H.~Eleuch and N.~Rachid, {\protect\JournalTitle{The European Physical Journal
  D}} \textbf{57}, 259 (2010).

\bibitem{Clark2017Sideband}
J.~B. Clark, F.~Lecocq, R.~W. Simmonds, J.~Aumentado, and J.~D. Teufel,
  {\protect\JournalTitle{Nature}} \textbf{541}, 191 (2017).

\bibitem{PhysRevLett.115.243603}
V.~Peano, H.~G.~L. Schwefel, C.~Marquardt, and F.~Marquardt,
  {\protect\JournalTitle{Phys. Rev. Lett.}} \textbf{115}, 243603 (2015).

\bibitem{PhysRevLett.113.151102}
Y.~Ma, S.~L. Danilishin, C.~Zhao, H.~Miao, W.~Z. Korth, Y.~Chen, R.~L. Ward,
  and D.~G. Blair, {\protect\JournalTitle{Phys. Rev. Lett.}} \textbf{113},
  151102 (2014).

\bibitem{1367-2630-19-8-083022}
W.-Z. Zhang, Y.~Han, B.~Xiong, and L.~Zhou, {\protect\JournalTitle{New Journal
  of Physics}} \textbf{19}, 083022 (2017).

\bibitem{PhysRevLett.118.143601}
M.~Korobko, L.~Kleybolte, S.~Ast, H.~Miao, Y.~Chen, and R.~Schnabel,
  {\protect\JournalTitle{Phys. Rev. Lett.}} \textbf{118}, 143601 (2017).

\bibitem{Liu:18}
S.~Liu, W.-X. Yang, Z.~Zhu, T.~Shui, and L.~Li, {\protect\JournalTitle{Opt.
  Lett.}} \textbf{43}, 9 (2018).

\bibitem{Liu:16}
N.~Liu, Y.~Liu, J.~Li, L.~Yang, and X.~Li, {\protect\JournalTitle{Opt.
  Express}} \textbf{24}, 2125 (2016).

\bibitem{Scully1998Quantum}
M.~O. Scully and M.~S. Zubairy, {\protect\JournalTitle{Cambridge University
  press}}  (1997).

\bibitem{1367-2630-18-7-073040}
A.~Motazedifard, F.~Bemani, M.~H. Naderi, R.~Roknizadeh, and D.~Vitali,
  {\protect\JournalTitle{New Journal of Physics}} \textbf{18}, 073040 (2016).

\bibitem{Abadie2011A}
J.~Abadie, B.~P. Abbott, R.~Abbott, T.~D. Abbott, M.~Abernathy, C.~Adams,
  R.~Adhikari, C.~Affeldt, B.~Allen, and G.~S. Allen,
  {\protect\JournalTitle{Nature Physics}} \textbf{7}, 962 (2011).

\bibitem{PhysRevLett.110.181101}
H.~Grote, K.~Danzmann, K.~L. Dooley, R.~Schnabel, J.~Slutsky, and H.~Vahlbruch,
  {\protect\JournalTitle{Phys. Rev. Lett.}} \textbf{110}, 181101 (2013).

\bibitem{PhysRevA.97.023841}
Y.-H. Chen, Z.-C. Shi, J.~Song, and Y.~Xia, {\protect\JournalTitle{Phys. Rev.
  A}} \textbf{97}, 023841 (2018).

\bibitem{PhysRevA.96.042320}
Q.~Zhang, X.~Zhang, and L.~Liu, {\protect\JournalTitle{Phys. Rev. A}}
  \textbf{96}, 042320 (2017).

\bibitem{PhysRevA.96.021801}
A.~Baksic, R.~Belyansky, H.~Ribeiro, and A.~A. Clerk,
  {\protect\JournalTitle{Phys. Rev. A}} \textbf{96}, 021801 (2017).

\bibitem{RevModPhys.86.1391}
M.~Aspelmeyer, T.~J. Kippenberg, and F.~Marquardt, {\protect\JournalTitle{Rev.
  Mod. Phys.}} \textbf{86}, 1391 (2014).

\bibitem{PhysRevLett.57.2520}
L.-A. Wu, H.~J. Kimble, J.~L. Hall, and H.~Wu, {\protect\JournalTitle{Phys.
  Rev. Lett.}} \textbf{57}, 2520 (1986).

\bibitem{PhysRevLett.76.2294}
X.~Hu and F.~Nori, {\protect\JournalTitle{Phys. Rev. Lett.}} \textbf{76}, 2294
  (1996).

\bibitem{doi:10.1021/nl101844r}
J.~Suh, M.~D. LaHaye, P.~M. Echternach, K.~C. Schwab, and M.~L. Roukes,
  {\protect\JournalTitle{Nano Letters}} \textbf{10}, 3990 (2010). PMID:
  20843059.

\bibitem{PhysRevLett.98.078103}
R.~Almog, S.~Zaitsev, O.~Shtempluck, and E.~Buks, {\protect\JournalTitle{Phys.
  Rev. Lett.}} \textbf{98}, 078103 (2007).

\bibitem{PhysRevA.91.013834}
X.-Y. L\"u, J.-Q. Liao, L.~Tian, and F.~Nori, {\protect\JournalTitle{Phys. Rev.
  A}} \textbf{91}, 013834 (2015).

\bibitem{Wang2016Steady2}
D.~Y. Wang, C.~H. Bai, H.~F. Wang, A.~D. Zhu, and S.~Zhang,
  {\protect\JournalTitle{Scientific Reports}} \textbf{6}, 38559 (2016).

\bibitem{PhysRevA.89.023849}
M.~Asjad, G.~S. Agarwal, M.~S. Kim, P.~Tombesi, G.~D. Giuseppe, and D.~Vitali,
  {\protect\JournalTitle{Phys. Rev. A}} \textbf{89}, 023849 (2014).

\bibitem{PhysRevA.89.063805}
M.~J. Woolley and A.~A. Clerk, {\protect\JournalTitle{Phys. Rev. A}}
  \textbf{89}, 063805 (2014).

\bibitem{PhysRevA.87.033829}
H.~Tan, G.~Li, and P.~Meystre, {\protect\JournalTitle{Phys. Rev. A}}
  \textbf{87}, 033829 (2013).

\bibitem{PhysRevA.88.063833}
A.~Kronwald, F.~Marquardt, and A.~A. Clerk, {\protect\JournalTitle{Phys. Rev.
  A}} \textbf{88}, 063833 (2013).

\bibitem{PhysRevA.83.033820}
J.-Q. Liao and C.~K. Law, {\protect\JournalTitle{Phys. Rev. A}} \textbf{83},
  033820 (2011).

\bibitem{Han2013}
Y.~Han, J.~Cheng, and L.~Zhou, {\protect\JournalTitle{The European Physical
  Journal D}} \textbf{67}, 20 (2013).

\bibitem{PhysRevA.91.022328}
J.~Cheng, W.-Z. Zhang, Y.~Han, and L.~Zhou, {\protect\JournalTitle{Phys. Rev.
  A}} \textbf{91}, 022328 (2015).

\bibitem{PhysRevA.94.012334}
Q.~Mu, X.~Zhao, and T.~Yu, {\protect\JournalTitle{Phys. Rev. A}} \textbf{94},
  012334 (2016).

\bibitem{PhysRevD.45.2843}
B.~L. Hu, J.~P. Paz, and Y.~Zhang, {\protect\JournalTitle{Phys. Rev. D}}
  \textbf{45}, 2843 (1992).

\bibitem{PhysRevA.91.012109}
S.~C. Hou, S.~L. Liang, and X.~X. Yi, {\protect\JournalTitle{Phys. Rev. A}}
  \textbf{91}, 012109 (2015).

\bibitem{PhysRevLett.109.170402}
W.-M. Zhang, P.-Y. Lo, H.-N. Xiong, M.~W.-Y. Tu, and F.~Nori,
  {\protect\JournalTitle{Phys. Rev. Lett.}} \textbf{109}, 170402 (2012).

\bibitem{RevModPhys.88.021002}
H.-P. Breuer, E.-M. Laine, J.~Piilo, and B.~Vacchini,
  {\protect\JournalTitle{Rev. Mod. Phys.}} \textbf{88}, 021002 (2016).

\bibitem{PhysRevA.96.062114}
H.-B. Chen, G.-Y. Chen, and Y.-N. Chen, {\protect\JournalTitle{Phys. Rev. A}}
  \textbf{96}, 062114 (2017).

\bibitem{PhysRevA.93.063853}
W.-Z. Zhang, J.~Cheng, W.-D. Li, and L.~Zhou, {\protect\JournalTitle{Phys. Rev.
  A}} \textbf{93}, 063853 (2016).

\bibitem{Shen:18}
H.~Z. Shen, S.~Xu, H.~Li, S.~L. Wu, and X.~X. Yi, {\protect\JournalTitle{Opt.
  Lett.}} \textbf{43}, 2852 (2018).

\bibitem{Valente:16}
D.~Valente, M.~F.~Z. Arruda, and T.~Werlang, {\protect\JournalTitle{Opt.
  Lett.}} \textbf{41}, 3126 (2016).

\bibitem{Li:17}
C.~Li, S.~Yang, J.~Song, Y.~Xia, and W.~Ding, {\protect\JournalTitle{Opt.
  Express}} \textbf{25}, 10961 (2017).

\bibitem{PhysRevA.97.012104}
J.~Jing, T.~Yu, C.-H. Lam, J.~Q. You, and L.-A. Wu,
  {\protect\JournalTitle{Phys. Rev. A}} \textbf{97}, 012104 (2018).

\bibitem{LI20182044}
C.~Li, J.~Song, Y.~Xia, and W.~Ding, {\protect\JournalTitle{Physics Letters A}}
  \textbf{382}, 2044  (2018).

\bibitem{1367-2630-20-5-053026}
J.~Jin and C.~shui Yu, {\protect\JournalTitle{New Journal of Physics}}
  \textbf{20}, 053026 (2018).

\bibitem{RevModPhys.89.015001}
I.~de~Vega and D.~Alonso, {\protect\JournalTitle{Rev. Mod. Phys.}} \textbf{89},
  015001 (2017).

\bibitem{groblacher2015observation}
S.~Gr{\"o}blacher, A.~Trubarov, N.~Prigge, G.~Cole, M.~Aspelmeyer, and
  J.~Eisert, {\protect\JournalTitle{Nat.Commun}} \textbf{6}, 7606 (2015).

\bibitem{RevModPhys.59.1}
A.~J. Leggett, S.~Chakravarty, A.~T. Dorsey, M.~P.~A. Fisher, A.~Garg, and
  W.~Zwerger, {\protect\JournalTitle{Rev. Mod. Phys.}} \textbf{59}, 1 (1987).

\bibitem{PhysRevA.81.052105}
K.~W. Chang and C.~K. Law, {\protect\JournalTitle{Phys. Rev. A}} \textbf{81},
  052105 (2010).

\bibitem{PhysRevLett.116.183602}
J.~F. Triana, A.~F. Estrada, and L.~A. Pach\'on, {\protect\JournalTitle{Phys.
  Rev. Lett.}} \textbf{116}, 183602 (2016).

\end{thebibliography}



\end{document}